\documentclass[12pt,a4paper]{article}
\usepackage{jheppub}
\pdfoutput=1

\usepackage{
graphicx,
subfig,
rotate,
color,
xcolor,
amsmath,
amssymb,
latexsym,
relsize,	
dcolumn,
bm,
epsfig
}

\begin{document}
%%%%%%%%%%%%%%%%%%%%%%%%%%%%%%%%%%%%%%%%%%%%%%%%%%%%%%%%%%%%%%%%%%%%%%%%%%%%%%%%%%%%%%%%%%
\newcommand{\yes}{\checkmark}
\newcommand{\nil}{$\times$}
\newcommand{\afb}{${\cal A}^t_{FB}$}
\hyphenation{FeynCalc}
\hyphenation{MadGraph}
\hyphenation{PYTHIA}
\def\lsim{\mathrel {\vcenter {\baselineskip 0pt \kern 0pt
    \hbox{$<$} \kern 0pt \hbox{$\sim$} }}}
\def\gsim{\mathrel {\vcenter {\baselineskip 0pt \kern 0pt
    \hbox{$>$} \kern 0pt \hbox{$\sim$} }}}

\catcode`\@=11
% \else
%\ProvidesPackage{slashed} [1997/01/16 v0.01 Feynman Slashed Character Notation (DPC)]
%\fi
\def\sla@#1#2#3#4#5{{%
 \setbox\z@\hbox{$\m@th#4#5$}%
 \setbox\tw@\hbox{$\m@th#4#1$}%
 \dimen4\wd\ifdim\wd\z@<\wd\tw@\tw@\else\z@\fi
 \dimen@\ht\tw@
 \advance\dimen@-\dp\tw@ \advance\dimen@-\ht\z@
 \advance\dimen@\dp\z@
 \divide\dimen@\tw@ \advance\dimen@-#3\ht\tw@
 \advance\dimen@-#3\dp\tw@ \dimen@ii#2\wd\z@
 \raise-\dimen@\hbox to\dimen4{%
 \hss\kern\dimen@ii\box\tw@\kern-\dimen@ii\hss}%
 \llap{\hbox to\dimen4{\hss\box\z@\hss}}}}
%%%%%%%%%%%%%%%%%%%%%%%%%%%%%%%%%%%%%%%%%%%%%%%%%%%%%%%%%%%%%%%%%%%%%%%%%%%%%%%%%%%%%%%%%%
\def\slashed#1{%
 \expandafter\ifx\csname sla@\string#1\endcsname\relax
{\mathpalette{\sla@/00}{#1}}
% \else \csname sla@\string#1\endcsname
\fi}
\def\declareslashed#1#2#3#4#5{%
 \expandafter\def\csname sla@\string#5\endcsname{%
#1{\mathpalette{\sla@{#2}{#3}{#4}}{#5}}}}
 \catcode`\@=12
\declareslashed{}{/}{.08}{0}{D}
 \declareslashed{}{/}{.1}{0}{A}
 \declareslashed{}{/}{0}{-.05}{k}
 \declareslashed{}{/}{.1}{0}{\partial}
 \declareslashed{}{\not}{-.6}{0}{f}
%%%%%%%%%%%%%%%%%%%%%%%%%%%%%%%%%%%%%%%%%%%%%%%%%%%%%%%%%%%%%%%%%%%%%%%%%%%%%%%%%%%%%%%%%%
%%%%%%%%%%%%%%%%%%%%%%%%%%%%%%%%%%%%%%%%%%%%%%%%%%%%%%%%%%%%%%%%%%%%%%%%%%%%%%%%%%%%%%%%%%
%%%%%%%%%%%%%%%%%%%%%%%%%%%%%%%%%%%%%%%%%%%%%%%%%%%%%%%%%%%%%%%%%%%%%%%%%%%%%%%%%%%%%%%%%%
\title{Constraining the flavor changing Higgs couplings to the top-quark at the LHC}
\author[a]{David Atwood,}
\author[b]{Sudhir Kumar Gupta,}
\author[c]{and Amarjit Soni}

\affiliation[a]{Dept. of Physics and Astronomy, Iowa State University,\\
 Ames, IA 50011, USA}
\affiliation[b]{ARC Centre of Excellence for Particle Physics at the Terascale,
School of Physics, Monash University, Melbourne, Victoria 3800 Australia}
\affiliation[c]{ Theory Group, Brookhaven National Laboratory, \\
Upton, NY, 11973, USA}

\emailAdd{atwood@iastate.edu}
\emailAdd{Sudhir.Gupta@monash.edu}
\emailAdd{adlersoni@gmail.com}
%%%%%%%%%%%%%%%%%%%%%%%%%%%%%%%%%%%%%%%%%%%%%%%%%%%%%%%%%%%%%%%%%%%%%%%%%%%%%%%%%%%%%%%%%%
\abstract
{
We study the flavor-changing couplings of the Higgs-boson with the 
top-quark using the processes: (a) $pp\longrightarrow tt$, (b) 
$pp\longrightarrow {\bar t} j$, and, (c) $pp\longrightarrow {\bar t} j h$ at the LHC 
in  light of current discovery of a 126 GeV Higgs-Boson. 
Sensitivities for the flavor-changing couplings  are  estimated 
 using the LHC data that was 
collected until spring 2013. It is  found that the process (c) is the most capable of 
yielding the best upper bound on the flavor-changing couplings with 
$2\sigma$ level sensitivities of ${|\xi_{tc}^2 + \xi_{tu}^2|}^{1/2} 
\lsim 4.2 \times 10^{-3}$ and $\lsim 1.7 \times 10^{-3}$ resulting from $t\to b l \nu_l, h \to jj$ with the 7 TeV
and 8 TeV centre-of-mass energies respectively using existing data from the LHC. The corresponding bounds from $h \to b {\bar b}$ are worse by a factor of about 1.8. 
%Our 14 TeV analysis for the $h\to jj$ case reveals that 
%the sensitivities can be further improved down to $1.1 \times 10^{-3}$ with 300 fb$^{-1}$ data.
}
%%%%%%%%%%%%%%%%%%%%%%%%%%%%%%%%%%%%%%%%%%%%%%%%%%%%%%%%%%%%%%%%%%%%%%%%%%%%%%%%%%%%%%%%%%

%\keywords{}
%\toccontinuoustrue
\arxivnumber{}
\maketitle
%\flushbottom
%%%%%%%%%%%%%%%%%%%%%%%%%%%%%%%%%%%%%%%%%%%%%%%%%%%%%%%%%%%%%%%%%%%%%%%%%%%%%%%%%%%%%%%%%%
%%%%%%%%%%%%%%%%%%%%%%%%%%%%%%%%%%%%%%%%%%%%%%%%%%%%%%%%%%%%%%%%%%%%%%%%%%%%%%%%%%%%%%%%%%

\section{\label{flh:int} Introduction}

Last year the ATLAS and the CMS experiments at the LHC made a monumental discovery, namely that of a scalar resonance with mass of about 126 GeV and properties akin to that of a Standard Model (SM) Higgs boson~\cite{atlas,cms}. By now all of the prominent decay modes that have been measured are found to be quite consistent within appreciable errors with expectations based on the SM~\cite{ATLAS-mass,CMS-mass}.

Given that this particle is of such a fundamental importance it should be clear that 
we need to study all its properties to excruciating details. In particular, we have to understand  the issue of its
stability against radiative corrections. Naively, one expects new physics  to be there around a few TeV scale. To 
decipher  its nature we are likely to need very precise measurements of the couplings of the Higgs.

In this work we will address the flavor changing couplings involving  the Higgs and the top-quark
and try to use data to
constrain these couplings. As is well known within the SM, such flavor off-diagonal couplings are highly suppressed and occur only at loop level. At least in some popular models, such as warped extra-dimension
the Higgs and the top-quark may be particularly sensitive to flavor changing effects as both of them are localized 
close to the TeV brane and the profiles of the KK-gluons are also peaked there ~\cite{Agashe:2004cp, Agashe:2006wa, Blanke:2008zb,  Blanke:2008yr, Casagrande:2008hr, Agashe:2009di, Azatov:2009na, Bauer:2009cf, KerenZur:2012fr}. We study several different
processes involving the top-quark and the Higgs and  find that
 $pp\longrightarrow {\bar t} j h$ , (where j is a jet), is very sensitive in providing stringent
constraints.

In warped models of course there are also general expectations  of  extra CP violating phases of ${\cal O}(1)$.
So CP violation studies are also highly motivated and we will return to this in another publication.

If such FCNCs are present, depending upon their size, they can influence the top-induced production processes, such as the single-top production processes where the top-quark is produced in 
association with lighter quarks, and Higgs, it can also modify (1) decay of the top-quark by allowing its decays into the Higgs-boson and a charm or an up-quark in addition to its known decays 
within the SM, and, (2) decay of the Higgs-boson  into a virtual top-quark in association with a charm or an up-quark where the virtual top-decays as  in (1).

The organisation of the paper is following: We begin with a brief introduction to the flavor-changing operators. In Sections 3 and 4 we will discuss decays of the Higgs boson and the top-quark, 
and the production processes we propose in this article. In section 5 we will present the actual analysis of the proposed processes in the light of LHC data on the Higgs-boson and the pre-existing 
data on the top-quark from the Tevatron experiment. Finally we summarise our findings in Section 6.

%%%%%%%%%%%%%%%%%%%%%%%%%%%%%%%%%%%%%%%%%%%%%%%%%%%%%%%%%%%%%%%%%%%%%%%%%%%%%%%%%%%%%%%%%%

\section{\label{flh:mod} The flavor-changing top-quark couplings}

When the flavor-changing couplings of the Higgs are present, the SM Lagrangian can be extended by allowing the  following additional terms,

\begin{eqnarray}
{\cal L}^h_{flavor} = \xi_{tc} {\bar t} c H +  \xi_{tu} {\bar t} u H  + h. c.,
\end{eqnarray}

where for now,  we consider that the flavor-changing couplings, $\xi_{tu}$, and, $\xi_{tc}$ are real and 
symmetric, i. e. $\xi_{tq_u} = \xi^\dagger_{tq_u} = \xi_{q_ut} = \xi^\dagger_{q_ut}$, where, $q_u 
= u, c$. As mentioned before, CP studies will be dealt with in a later study.

It is to be noted here that similar flavor-changing Lagrangian has been also widely studied in the Refs.~\cite{Cheng:1987rs, Luke:1993cy, Atwood:1996vj, Aquino:2006vp, Craig:2012vj, Chen:2013qta,Zhang:2013xya}, for example.
These FV-interactions allow some interesting phenomenological consequences which are  as follows:

{\em (A) At the decay level}

1. In addition to the usual decay modes, Higgs can also decay into single $W^\pm$-boson via an 
off-shell top, e.g. $h\to {\bar q}_u (t^\ast \to b W^+)$; $(q_u = u, c)$ where `$^\ast $` means off-shell top,

2. top-quark can now also decay into a charm or an up-quark and a Higgs, e.g. $t\to q_u h$.

{\em (B) At the production level}

1. Because of FCNC we can have a pair of same-sign top produced via t-channel exchange of the Higgs, e.g. $pp\to t t ({\bar t} {\bar t})$ through the parton level subprocesses, $uu\to t t $, $u c \to t t $ and $c c \to t t $.

2. The FCNCs of Higgs can contribute significantly to the production processes where a top-quark is produced in association with light partons, e.g. $p p \to t {\bar j}_u ({\bar t} j_u)$, where $j_u = u, c$.

3. Higgs can now be produced in association with a top-quark and an up or a charm quark, e.g. the process $pp\to t \bar {c} h ( {\bar t} c h)$.

Let us discuss their consequences one-by-one in the following sections.

\section{\label{flh:dec} Higgs boson and top-quark decay rates}
Let us now consider the decays of top-quark and Higgs-boson in the following subsections:

%%%
\subsection{\label{flh:tdecay} Top quark decays}
Because $m_t - m_h > m_c, m_u, m_b$, in addition to the usual decay into a bottom quark and a $W^+$ Boson, the top-quark can also decay into a charm (or an up) quark and a Higgs-Boson. 
Therefore, the total decay width of the top-quark, $\Gamma_t$ will take the following form, 
\begin{eqnarray}
\Gamma_t &=&\Gamma _{t\to bW^+}+\Gamma _{t\to c h}+\Gamma _{t\to u h}.
\end{eqnarray}

The $t\longrightarrow b W^+$ decay width at the NLO level is given by~\cite{Jezabek:1988iv, Beringer:1900zz},

\begin{eqnarray}
\Gamma_{t\to bW^+} &=&\frac{G_F m_t^3}{8\sqrt{2} \pi} \left[1-\frac{2 \alpha_s}{3 \pi }\left(\frac{2 \pi ^2}{3}-\frac{5}{2}\right)
   \right] \left(1-\tau _W^2\right){}^2 \left(2 \tau _W^2+1\right){}
\end{eqnarray}

The $t\longrightarrow q_u h$ ($q_u = u, c$) partial decay width is given as, 

\begin{eqnarray}
\Gamma_{t\to q_u h} &=& \frac{\xi _{t{q_u}}^2 m_t}{16 \pi }  \left[\left(\tau _{q_u}+1\right){}^2-\tau _h^2\right]
   \sqrt{1-\left(\tau _h-\tau _{q_u}\right){}^2} \sqrt{1-\left(\tau _{q_u}+\tau
   _h\right){}^2}
\end{eqnarray}

where, $\tau_{W} = \frac{M_W}{m_t}$, $\tau_{h} = \frac{M_h}{m_t}$, $\tau_{q_h} = \frac{m_{q_h}}{m_t}$. Using the measured values of $G_F$, $\alpha_s$, $M_W$, $M_h$ etc., we obtain,

\begin{eqnarray}
\Gamma_t &=& \Gamma_t^{SM} + [0.8~\xi_{tc}^2 + 0.78~\xi_{tu}^2] {~\rm GeV}, 
%{\cal B}_{t\to q_u h} &=&
\end{eqnarray}

where $\Gamma_t^{SM} =2 \pm 0.7$ GeV~\cite{Beringer:1900zz} and $\Gamma_t^{SM} \equiv \Gamma_{t \to b W^+}^{SM} $.

From the above expression it is clear that the 
maximum deviation in the $Br(t\to b W^+)$ from the SM corresponds to the $\xi_{tc} = 1= \xi_{tu}$ and amounts to be about $41\%$. As we will show later in this paper, the current LHC data constraints on 
these FCNCs will be much stronger, i. e. $\lsim 10^{-3}$. To put things into perspective, in the SM, the diagonal couplings of the top-quark and the charm-quark 
 $y_{t, c} = \frac{\sqrt{2}m_{t, c}} {v}$; $v = 246$ GeV being the vaccum-expectation value, are $y_t \simeq 1$ and  $y_c \simeq 0.008$ respectively.

\subsection{\label{flh:hdecay} Decays of the Higgs-Boson}
Because of the flavor-changing couplings endowed  by Eqn. (2.1) the Higgs-Boson can also have the following three-body decays:

$h\to q_u ({\bar t}^\ast \to {\bar b} W^-)$, $h\to {\bar q}_u (t^\ast \to b W^+)$; $(q_u = u, c)$ where `$^\ast $` means off-shell top. Using {\tt CalcHep}~\cite{Belyaev:2012qa} we numerically estimate, 

\begin{eqnarray}
\Gamma_{h\to q_u ({\bar t}^\ast \to {\bar b} W^-)} &\simeq& 0.28 \xi^2_{tq_u} {~\rm MeV}.
\end{eqnarray}
Therefore the total width, $\Gamma_h$ would be, 
\begin{eqnarray}
\Gamma_h &=& \Gamma^{SM}_h 
+ \sum\limits_{q_u}\Gamma_{h\to q_u ({\bar t}^\ast \to {\bar b} W^-)}  
+  \Gamma_{h\to {\bar q}_u (t^\ast \to b W^+)}\nonumber\\
&\simeq& \left[\Gamma^{SM}_h + 0.56 \left( \xi^2_{tc} +  \xi^2_{tu}\right) \right]  {~\rm MeV}, 
\end{eqnarray}
where $\Gamma^{SM}_h = 3.3$ using {\tt CalcHep}.

In our model the usual two-body decays of the Higgs boson, i.e.,  the LHC Higgs observables would take the following form, 
\begin{eqnarray}
{\cal R}_{ggX} &=&  \frac{\sigma_{gg\to h}}{\sigma^{SM}_{gg\to h}} \cdot \frac{{\cal BR}(h\to 
X)}{{\cal BR}^{SM}(h\to X)}
\simeq \frac{\Gamma_{h\to gg}}{\Gamma_{h\to gg}^{SM}} \cdot 
{\left(\frac{\Gamma_{h\to X}}{{\Gamma_h}}\right)}/{\left(\frac{\Gamma^{SM}_{h\to X}}{{\Gamma^{SM}_h}}\right)} 
\simeq \frac{\Gamma^{SM}_h}{\Gamma_h} ;
\end{eqnarray}
where, $X$ stands for two-body decay products of the produced Higgs-Boson, e.g. 
$X = \gamma\gamma, b\bar{b}, WW^\ast,  ZZ^\ast, \tau^+\tau^-$ etc. In the equation above, we have assumed that $\frac{\sigma_{gg\to h}}{\sigma^{SM}_{gg\to h}} = \frac{\Gamma_{gg\to h}}{\Gamma^{SM}_{gg\to h}} \simeq 1$ which is due to the fact that the 
flavor-changing couplings discussed in our paper affect the $gg\to h$ production process at the next-loop level and therefore the effects are expected to be much smaller.

%%%%%%%%%%%%%%%%%%%%%%%%%%%%%%%%%%%%%%%%%%%%%%%%%%%%%%%%%%%%%%%%%%%%%%%%%%%%%%%%%%%%%%%%%%

\section{\label{flh:prod} Production Processes at the LHC}
%%%

We now turn to study some interesting production processes where the 
effect of the aforementioned flavor-changing couplings could be 
significant. Obviously the very first process, the $t\bar{t}$ may not be suitable when both the produced tops decay through their ususal SM decay model as it will have relatively large SM background. Among 
the other leading processes are: 
\itemize
\item (a) the same-sign top-pair, e.g. $pp\to t t ( {\bar t} {\bar t})$, 
\item (b) processes where a top-quark is produced in association with a light 
jet, e.g. $p p \to t {\bar j}_u ({\bar t} j_u)$, where $j_u = u, c$, and, 
\item (c) processes where the top-quark is produced in association 
with a Higgs and a light jet, e.g. $p p \to t {\bar j}_u h ({\bar t} j_u  h)$. 

The process (a), the same-sign top pair production is an interesting 
one as it has very little SM background. In a recent work by 
us~\cite{Atwood:2013xg}, it was found that this process can play a 
very important role in excluding many models proposed to explain the 
forward-backward asymmetry of the top-quark $A^t_{FB}$ as observed by 
the Tevatron experiments (Also see, for example, ~\cite{Berger:2011ua, Degrande:2011rt, AguilarSaavedra:2011zy}). Within the model under consideration in this 
paper, at the LHC, a pair of same-sign top pair can be produced via the 
t-channel exchange of the Higgs through two flavor-changing couplings. 
For the $m_h = 125.7$ GeV, the bare cross-sections for this process is 
given by the following equations:

\begin{eqnarray}
\sigma(pp \to tt) &=& 49.5 \xi_{tu}^4 + 13.6 \xi_{tc}^2 \xi_{tu}^2 + 0.3 \xi_{tc}^4  {~\rm pb~(at~\sqrt{s} = 7~ TeV),}\nonumber\\
\sigma(pp \to tt) &=& 56.2 \xi_{tu}^4 + 17.7 \xi_{tc}^2 \xi_{tu}^2 + 0.4 \xi_{tc}^4  {~\rm pb~(at~\sqrt{s} = 8~ TeV),} 
\label{xtt}
\end{eqnarray}
respectively.

In Eqns. above we notice that the numerical factors for the 
$\xi_{tu}^4$ is larger than those proportional to $\xi_{tc}^2 \xi_{tu}^2$ 
and $\xi_{tc}^4$. This is because of the fact that the first term is 
due to the scattering of two valence up-quarks as opposed to other terms 
where one or both the initial partons are the (sea-)charm-quark(s). As 
discussed in Refs.~\cite{Atwood:2013xg} and in~\cite{Gupta:2010wx} this 
process has very little SM background and hence can be useful in 
constraining the couplings $\xi_{tc}$ and $\xi_{tu}$. We will discuss this in 
detail in a later section.

The other two processes namely, (b) and (c) are perhaps even more interesting as both 
of these occur via only one flavor-changing vertex unlike the process 
(a). Therefore, we expect the cross-sections for these processes to be 
proportional to $a \xi_{tu}^2 + b \xi_{tc}^2$ where a and b are arbitrary 
constants.

More specifically, the processes where a top or anti-top quark is produced in association with an up, anti-up, charm or anti-charm quark can occur via two ways: one, where a pair of initial 
(anti-)quarks with same or different flavor undergo t-channel exchange through the Higgs, e.g. processes of the form $q q_u \to t q$ and $q {\bar q}_u \to {\bar t} q$, where, $q =\{q_u, b\}$, 
$q_u = \{u, c\}$. Therefore we expect cross-sections for these subprocesses to be proportional to $y_q^2 \xi_{tq_u}^2$ where $y_q$ is the Yukawa coupling of the quark q. This 
suggests that although at the subprocess level, we gain in the processes where the initial parton is a b or a ${\bar b}$ due to large value of $y_b$ compared to $y_c$ and $y_u$ at the proton-proton 
level, we may loose due to relatively smaller parton densities for the b-quarks. The other type of contribution is due to the off-shell Higgs production in the s-channel due to gluon-pair fusion, 
where the Higgs decays into $t \bar {q}_u$ or ${{\bar t} q_u}$. The composition of this process is discussed in detail for $m_h = 125.7$ GeV in the Appendix A. The corresponding background and the 
total cross-sections for $\sqrt{s} = 7$ and 8 TeV are summarised in Table~\ref{t:sigma_22}. The major source for this process is production of $t{\bar b} ({\bar t} b)$ mediated by an off-shell $W^\pm$ in the s-channel.

\begin{figure}
\centerline{\includegraphics[angle=-00, width=1.\textwidth]{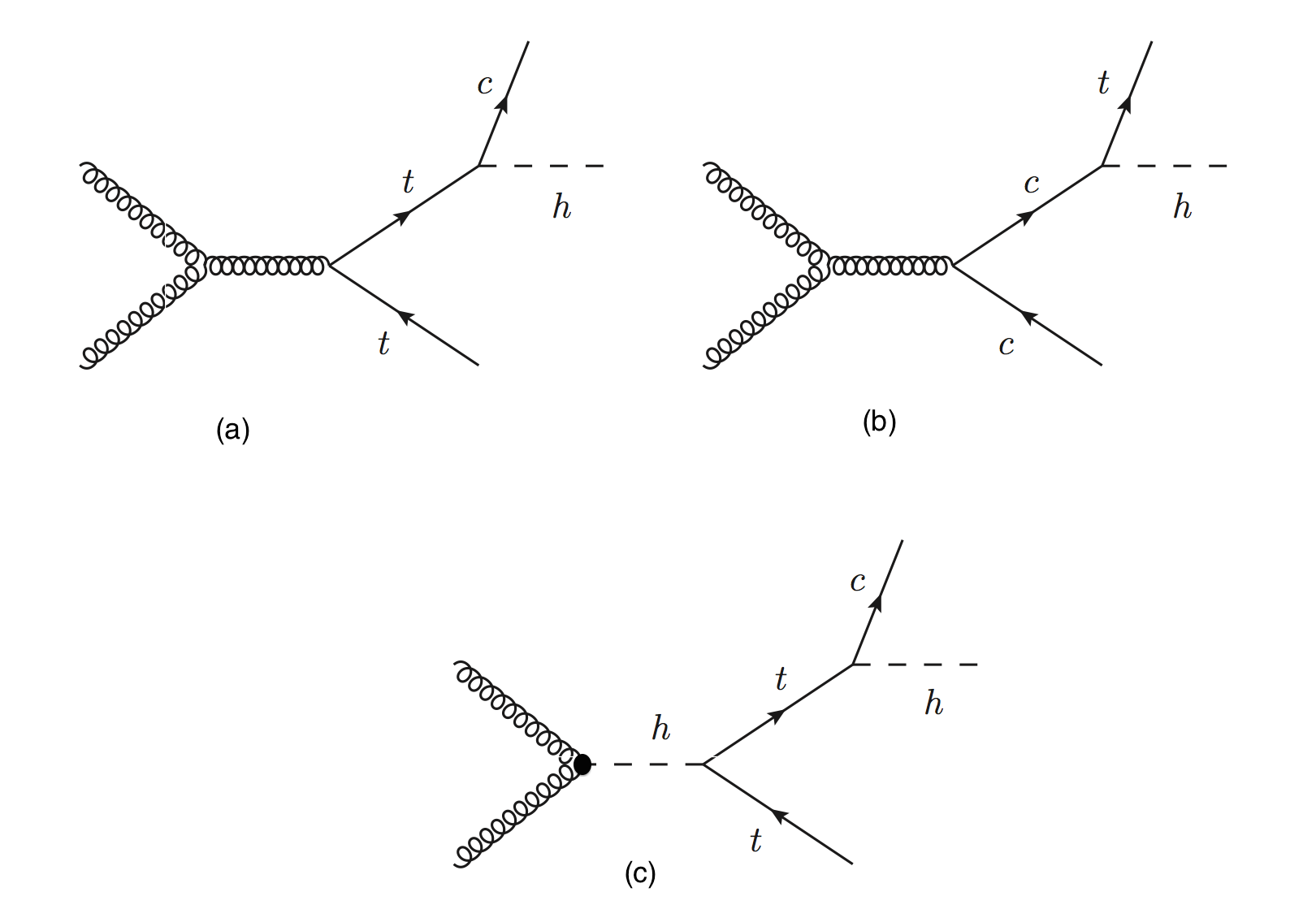}}
\caption{\sf\small Representative s-channel Feynman diagrams for the partonic process $g g\to t c h$.}
\label{fig:fg}
\end{figure}

The production process (c) requires production of a $t{\bar t}$ pair 
where one of the top decays into a Higgs and an up or charm-quark, or it can also take place via the pair production processes 
$u{\bar u}$ and $c\bar{c}$ where one of the off-shell up or charm goes into a top and a Higgs (see Fig~\ref{fig:fg}). This process has very little SM background due to the production of (i) $t 
{\bar b} Z ({\bar t} b Z)$, and, (ii) $t {\bar b} h ({\bar t} b h)$. 
Clearly both of these processes are electro-weak processes which are 
mediated by an off-shell $W^\pm$-boson. e.g., $pp\to {W^\pm}^\ast Z 
({W^\pm}^\ast h)$, and ${W}^\ast \to {t \bar b} ({\bar t b})$. Therefore 
we expect to find better bounds on $\xi_{tc}$ and $\xi_{tu}$ using this 
process. A detail of the cross-sections and corresponding background to 
this process is given in Table~\ref{t:sigma_23}.

%----------------------------------------------------------------------------------------%
%----------------------------------------------------------------------------------------%

%----------------------------------------------------------------------------------------%
\begin{table}
\centering
\resizebox{0.9\hsize}{!}
{
\begin{tabular}{|c|c|c|c|}\hline\hline
{\bf Serial Number} &{\bf Process} & {\bf LHC-7}& {\bf  LHC-8}\\\hline
(1)&$pp\longrightarrow {t}{\bar c} ({\bar t }{ c}, tc, {\bar t } {\bar c})$ & $42.06\xi^2_{tc} +4.81\xi^2_{tu}$ & $59.7\xi^2_{tc} +5.44\xi^2_{tu}$ \\ 
(2)&$pp\longrightarrow h \longrightarrow {t}{\bar c} ({\bar t }{ c})$  & $41.2\xi^2_{tc}$ & $58.52\xi^2_{tc}$ \\ \hline
(3)&$pp\longrightarrow {t}{\bar u} ({\bar t }{ u}, tu, {\bar t } {\bar u})$ &$ 0.4\xi^2_{tc} +44.54\xi^2_{tu}$ & $ 0.61\xi^2_{tc} +59.67\xi^2_{tu}$ \\ 
(4)&$pp\longrightarrow h \longrightarrow {t}{\bar u} ({\bar t }{ u})$ & $41.2\xi^2_{tu}$&$58.52\xi^2_{tu}$ \\ \hline
(1) + (3) &$pp\longrightarrow {t}{q_u} ({\bar t }{ q_u})$ & $42.46\xi^2_{tc} +49.35\xi^2_{tu}$& $60.31\xi^2_{tc} +65.11\xi^2_{tu}$ \\
\hline
SM Background &$pp \longrightarrow t {\bar b} (t {\bar b}, t j, {\bar t} j, t jj, {\bar t} j j)$  & $ 43.49 \times 10^3$ & $56.17 \times 10^3$  \\\hline\hline
\end{tabular}
}
\caption{\sf\small Single top production cross-section (in fb units) at the LHC for $\sqrt{s} = 7$ and 8 TeV in association with an up or a charm (anti-)quark. In all the processes the implemented basic 
cuts on the associated (anti-)quark are as follows: $p_T = 25$ GeV, and, $|\eta| \leq 2.7$. The Yuwaka couplings $y_u$, $y_c$, $y_t$ have been estimated using $m_u = 3$ MeV, $m_c = 1.44$ GeV and 
$m_t = 172.4$ GeV respectively. The Higgs-Boson mass has been set to the observed central value, $m_h = 125.7$ GeV.}
\label{t:sigma_22}
\end{table}
%----------------------------------------------------------------------------------------%

%----------------------------------------------------------------------------------------%
\begin{table}
\centering
\resizebox{1.0\hsize}{!}
{
\begin{tabular}{|c|c|c|c|}\hline\hline
{\bf Serial Number} &{\bf Process} & {\bf LHC-7}& {\bf  LHC-8}\\\hline
(5)&$pp\longrightarrow {t}{\bar c} h ({\bar t }{c} h, tc h, {\bar t } {\bar c} h)$ &$75.25\times 10^3 \xi^2_{tc}$& $107.99\times 10^3 \xi^2_{tc}$\\
(6)&$pp\longrightarrow {t}{\bar u} h ( {\bar t }{u} h, tu h, {\bar t } {\bar u} h)$ & $79.21\times 10^3 \xi^2_{tu}$& $112.76\times 10^3 \xi^2_{tu}$ \\ \hline
(5) + (6) &$pp\longrightarrow {t}{\bar q}_uh ({\bar t }{q_u}h)$ & $(75.25\xi^2_{tc} + 79.21\xi^2_{tu})\times 10^3$& $(107.99\xi^2_{tc} +112.76\xi^2_{tu})\times 10^3$ \\\hline
SM Background &$pp \longrightarrow t {\bar b} Z ({\bar t} b Z)$  & 2.54 (1.07)& 3.32 (1.39)  \\
              &$pp \longrightarrow t {\bar t} + n j; n \leq 3$ & $176.6 \times 10^3$ & $261.5\times 10^3$\\
              &$pp \longrightarrow t {\bar t} + b{\bar b} + n j; n \leq 3$ & $716.9$ & $1175.6$\\
&$pp \longrightarrow t {\bar b} h ({\bar t} b h)$  & 0.54 (0.23)& 0.7 (0.29) \\\hline
\end{tabular}
}
\caption{\sf\small Single top production cross-section (in fb units) at the LHC for $\sqrt{s} = 7$ and 8 TeV in association with an up or a charm (anti-)quark and a Higgs-Boson. In all the processes the 
implemented basic cuts and all other SM parameters are same as given in Table~\ref{t:sigma_22}.}
\label{t:sigma_23}
\end{table}
%----------------------------------------------------------------------------------------%

%%%%%%%%%%%%%%%%%%
\begin{table}
\centering
%\resizebox{12.5cm}{!}
{
\begin{tabular}{|c|c|c|}\hline\hline
{\bf Observable} & {\bf Value}& {\bf Experiment}\\\hline
$\Gamma_t$    &  $2\pm 0.7$ GeV &{\tt Tevatron}~\cite{Beringer:1900zz}\\\hline
$m_h$                          &  $125.5\pm 0.2~{\rm (stat)}^{+0.5}_{-0.6}~{\rm (sys)}$ GeV & {\tt ATLAS}~\cite{ATLAS-mass}\\
                               &  $125.8 \pm 0.5~{\rm (stat)} \pm 0.2~{\rm (sys)}$ GeV& {\tt CMS}~\cite{CMS-mass}\\
                               &  $125.7 \pm 0.4$ GeV & {\tt Combined}     \\\hline
${\cal R}_{gg\gamma\gamma}$    &  $1.65^{+0.34}_{-0.30}$ & {\tt ATLAS}~\cite{ATLAS-gamma, ATLAS-bosonic}\\ 
                               &  $1.11^{+0.32}_{-0.30}$ & {\tt CMS}~\cite{CMS-bosonic-QCD, CMS-gamma}\\
                               &  $1.36 \pm 0.23$ & {\tt Combined}     \\\hline
${\cal R}_{gg2l2\nu}$          &  $1.01\pm0.31$ & {\tt ATLAS}~\cite{ATLAS-bosonic-QCD, ATLAS-WW}\\
                               &  $0.76^{+0.21}_{-0.21}$ & {\tt CMS}~\cite{CMS-bosonic, CMS-WW}\\
                               &  $0.84 \pm 0.17$ & {\tt Combined}     \\\hline
${\cal R}_{gg4l}$              &  $1.7^{+0.5}_{-0.4}$ & {\tt ATLAS}~\cite{ATLAS-ZZ,ATLAS-bosonic}\\
                               &  $0.91^{+0.30}_{-0.24}$ & {\tt CMS}~\cite{CMS-bosonic, CMS-ZZ}\\
                               &  $1.12 \pm 0.26$ &{\tt Combined}     \\\hline
%${\cal R}_{ggbb}$              &  $1.09\pm0.20\pm0.22$ & {\tt ATLAS}~\cite{ATLAS-fermionic, ATLAS-bb}\\
%                               &  $1.3^{+0.7}_{-0.6} $ & {\tt CMS}~\cite{LHC-fermionic} \\
%                               &  $1.12 \pm 0.27$ &{\tt Combined}     \\\hline
%${\cal R}_{gg\tau\tau}$        &  $0.7\pm 0.7$ & {\tt ATLAS}~\cite{LHC-fermionic, ATLAS-tautau}\\
%                               &  $1.1^{+0.4}_{-0.4}$ & {\tt CMS}~\cite{CMS-fermionic, CMS-tautau}\\
%                               &  $1.0 \pm 0.35$ &{\tt Combined}     \\\hline
$D0$-oscillations              & $\left|\xi_{tu} \xi_{tc} \right| < 0.9 \times 10^{-3}$& 
{\tt UTfit Collaboration}~\cite{Bona:2007vi}\\\hline
\hline
\end{tabular}
}
\caption{\sf Measured values of various observables used in our analysis; combined here means weighted average of ATLAS and CMS values for a given observable.}
\label{t:obs}
\end{table}
%%%%%%%%%%%%%%%%%%%

\section{\label{flh:analysis} Flavor-changing Higgs and LHC observations}

In this Section we will first discuss the flavor-changing couplings $\xi_{tc}$ and $\xi_{tu}$ in the 
context of the individual experimental observations as listed in Table~\ref{t:obs} and later we will 
make use of them all in order to obtain constraints on these couplings using the processes we propose 
here. We will also discuss the projected sensitivities of the aforementioned couplings in the context of 
14 TeV data in the second Subsection.

\subsection{Flavor changing couplings at the LHC-7 and LHC-8}

Let us first begin with total decay width of the top-quark, $\Gamma_t$. 
As we have already noticed in Section~\ref{flh:tdecay}, $\Gamma_t$ 
receives positive contributions proportional to $\xi_{tu}^2$ and to 
$\xi_{tc}^2$ due to additional decay processes $t\to u h$ and $t\to c h$ 
respectively. 

This gives an upper bound on the 
$\sqrt{\xi_{tc}^2 + \xi_{tu}^2}$ of about $1.3$ at the $2\sigma$ level which 
is quite mild.

%----------------------------------------------------------------------------------------%
\begin{figure}
\centerline{\includegraphics[angle=-00, width=1\textwidth]{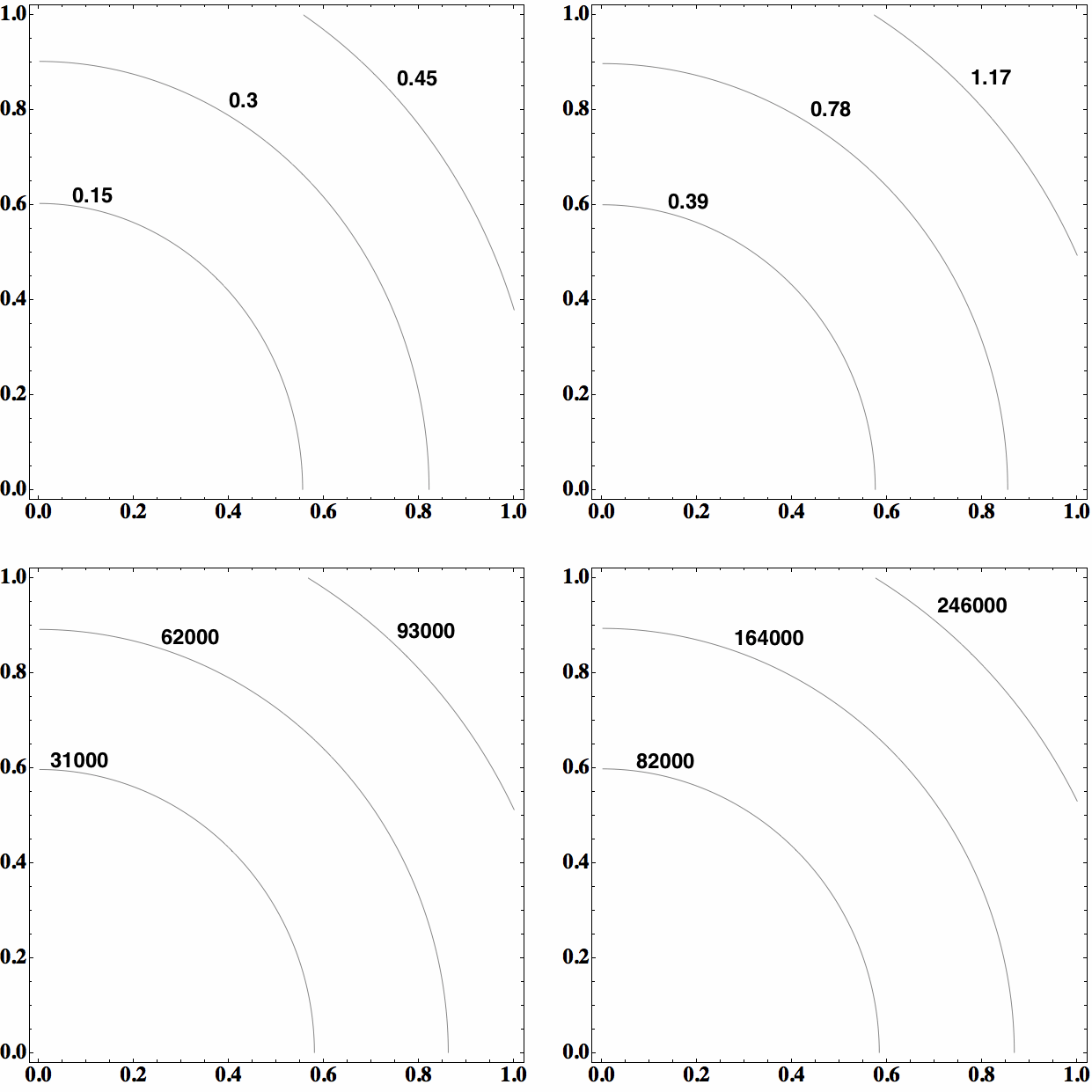}}
\caption{\sf\small Contour plots in $\xi_{tc} - \xi_{tu}$ plane for signal-significance, $\frac{S}{\sqrt{B}}$, at the production level, for the processes $pp\longrightarrow t {\bar j}_u ({\bar t} j_u)$ (top) and 
$pp\longrightarrow t {\bar j}_u h ({\bar t} j_u h)$ (bottom) 
for $\sqrt{s} = 7$ (left) and $8$ TeV (right) data at the LHC.}
\label{fig:sprod}
\end{figure}
%----------------------------------------------------------------------------------------%

In a similar way we study the Higgs observables ${\cal R}_{gg\gamma\gamma}$, ${\cal R}_{gg 2l2\nu}$, ${\cal R}_{gg4l}$ 
and their combined effects in light of recently updated data during 
Moriond-2013 as discussed in Table~\ref{t:obs}. 
%We plot contour for each 
%of them in Figs.~\ref{fig:stdhiggs} at various $sigma$-levels using the combined ATLAS and CMS results. 
 Using the LHC data on various Higgs-observables, one obtains upper 
bounds on $\sqrt{\xi_{tc}^2 + \xi_{tu}^2}$ of about $1.7$, $3.3$, $1.8$, 
and, $1.2$, for ${\cal R}_{gg\gamma\gamma}$, ${\cal R}_{gg2l2\nu}$, ${\cal R}_{gg4l}$ and for their combined effect 
respectively. This means the whole range of $\xi_{tc}$ 
and $\xi_{tu}$ (between -1 to +1) is allowed by these observables at the $2\sigma$ level. Note also that the bound that we will obtain later in this paper from $pp\to tch$ will be much stronger,  $\lsim 10^{-3}$.

In order to study the production processes we incorporated flavor-changing couplings in {\tt MadGraph5}~\cite{madgraph}. We evaluate parton densities at a scale $\mu_R = \sqrt{\hat s} =\mu_F$ using 
{\tt CTEQ6L1}~\cite{cteq}.

We present our result for the $\sqrt{s} = 7$ TeV with an integrated luminosity of $\int {\cal L} dt = 5$ fb$^{-1}$ and for $\sqrt{s} = 8$ TeV with  $\int {\cal L} dt = 22$ fb$^{-1}$ in Figures~\ref{fig:sprod} at the bare production level.

Using the formulae for the production cross-sections in our model and their respective SM backgrounds as mentioned in Eqns.~\ref{xtt} and Tables~\ref{t:sigma_22} and~\ref{t:sigma_23}, and the constraints due to the $\Gamma_t$ and the LHC Higgs discovery observables $R_{gg\gamma\gamma}$, $R_{gg2l2\nu}$, and, $R_{gg4l}$,  we find the following upper bounds on the $\sqrt{\xi_{tc}^2 + \xi_{tu}^2}$ at the 2$\sigma$ level;

\begin{eqnarray}
\sqrt{\xi_{tc}^2 + \xi_{tu}^2} &\lsim& \begin{cases} 0.3~(0.17) & {\rm for~process~(a)} \\
0.9~(0.9) & {\rm for~process~(b)} \\
1.6~(0.6)\times 10^{-3} & {\rm for~process~(c)}\end{cases}.
\end{eqnarray}
for the $\sqrt{s} =$ 7 TeV with 5 fb$^{-1}$ ($\sqrt{s} =$ 8 TeV with 22 fb$^{-1}$) data.

Clearly the bounds due to process (b) are not so promising which is partially due to the 
fact that it suffers from large SM background as mentioned in Table~\ref{t:sigma_22}. The reason for them being the same for 
both the 7 TeV and 8 TeV LHC centre-of-mass energies is that it in determining these the combined 
effect of the constraints on the top-quark decay width and the Higgs observables play a dominant role. Note 
also that although the processes (a) puts mild bound on $\sqrt{\xi_{tc}^2 + \xi_{tu}^2}$ it is 
interesting to analyse because it leads to unique signature in the form of a pair of 
same-sign leptons. The process (c) is certainly the best process among all of the 
aforementioned ones as it turns out to be very sensitive even to small values of the 
flavor-changing couplings.

It is to be noted that in order to obtain the aforementioned bounds, we 
have assumed that, (1) we will be able to reconstruct the produced 
(anti-)top-quark(s), and the Higgs boson fully in all its detection 
modes, and, (2) the number of signal events do not exceed one where the
SM background is not significant for the given LHC luminosity. However due to poor reconstruction especially for the cases where 
the top-quark(s) decays hadronically into a b-jet and a pair of light 
parton jet, the above limits may not be so realistic. We therefore turn 
our focus to obtain the detection level bounds on these couplings. For 
this we work with processes with at least one lepton in the final state.  
Thus, in all the production processes we allow the produced 
(anti)top-quark to decay semileptonically, e.g, $t\to b l \nu_l$, where 
$l = e, \mu$. In case of the process (c) where a Higgs is also produced 
in association with the top-quark and a jet, we work with $h \to b {\bar 
b}$, $h\to \gamma\gamma$ and $h\to j j$, where $j = g, q, \bar q, b, \bar 
b$. The 
reason for considering $h\to b {\bar b}$ is merely to gain statistical 
advantage as within SM the branching ratio for $h\to b {\bar b}$ for the 
given value of $m_h$ is about $79\%$. For the case where the Higgs decays 
into a pair of photons, although the branching ratio is quite suppressed 
$\simeq 2.9 \times 10^{-3}$, we expect it to be relatively cleaner than 
the $h\to b{\bar b}$ case.

With this in mind we will therefore have the following topologies: 
\begin{itemize}
\item $l^\pm l ^\pm + 2b-jets + {\slashed E}_T$, from the process (a), 
\item $l^\pm  + j + b-jet + {\slashed E}_T$, from process (b), and, 
\item $l^\pm  + j + 3b-jets +  {\slashed E}_T$, when $h\to b{\bar b}$, $l^\pm  + j + b-jet + 2\gamma + {\slashed E}_T$, when $h\to \gamma\gamma$, and, 
$l^\pm  + 3j + b-jet +  {\slashed E}_T$, when $h\to j j$, from process (c).
\end{itemize}

The set of basic cuts used on the photons/leptons/jets in our study are as follows: 

$p_{T_{l,j,\gamma}} > 25$ GeV, $\left|\eta_{l,j,\gamma}\right| < 2.7$, $\Delta R_{kk}, \Delta R_{ik}, > 0.4$, $\slashed E_T > 30$ GeV, with $i, k=\{l, j, \gamma\}$. In addition, we also assume a 
b-jet identification efficiency of $58\%$~\cite{btag}. 

{\underline {Other SM Backgrounds:}} It is to be noted that because a light parton jet can fake the b-jet with probabilities of about $10\%$ for a charm-jet and about $1\%$ for other light jets respectively~\cite{fakejet}), 
there can be other subleading SM backgrounds for the process (c), 
particularly when $h\to b \bar {b}$ and $t\to b l \nu_l$. In this catagory, the leading 
background contribution come from the processes $pp\to t\bar{t} + n-jets$ 
and $pp\to t\bar{t} + b \bar{b} + n-jets$, where n represents number of 
jets.{\footnote{We thank J. Evans for discussion on this background.}

Using {\tt MadGraph}, for $n \leq 3$, cross-sections for these processes 
have been estimated to be $\simeq 177$ pb and $\simeq 0.7$ pb respectively at the 
$\sqrt{s} = 7$ TeV at the bare level. With the above sets of basic cuts, 
for our final state $pp \to l^\pm + j + 3 b-jets + {\slashed E}_T$, the 
requirement that one of the jet pair reconstructs to $m_h$ within $2\sigma$, these 
aforementioned backgrounds are reduced to $0.33$ fb and $0.0014$ fb 
respectively. The corresponding signal rates are reduced by 0.42, 0.57 and 0.43 respectively for the processes $l^\pm  + j + 3b-jets +  {\slashed E}_T$, $l^\pm  + j + b-jet + 2\gamma + {\slashed E}_T$, and, $l^\pm  + 3j + b-jet +  {\slashed E}_T$. Similar estimates for $\sqrt{s} = 8$ TeV have been summarised in Table~\ref{t:bkgother} and \ref{t:rbkgother}. Thus our estimates suggest that although at the production level such 
backgrounds are huge for the LHC integrated luminosities we consider in our analysis, with the aforementioned cuts, these can be reduced to a level where their effects become insignificant for our purposes.}

%----------------------------------------------------------------------------------------%
\begin{table}
\centering
\resizebox{0.5\hsize}{!}
{
\begin{tabular}{|c|c|c|}\hline\hline
{\bf Process} & {\bf LHC-7}& {\bf  LHC-8}\\\hline
$pp\longrightarrow {t}{\bar t} + n-jets$ & 176.64 & 261.52 \\
$pp\longrightarrow {t}{\bar t} + b\bar{b} +  n-jets$ & 0.72  & 1.18 \\\hline
Total   &     177.36 &      262.7 \\\hline\hline
\end{tabular}
}
\caption{\sf\small Other subleading SM backgrounds (in pb) for the production process (c) for $\sqrt{s} = 7$ and 8 TeV.}
\label{t:bkgother}
\end{table}
%----------------------------------------------------------------------------------------%

%----------------------------------------------------------------------------------------%
\begin{table}
\centering
\resizebox{0.5\hsize}{!}
{
\begin{tabular}{|c|c|c|}\hline\hline
{\bf Process} & {\bf LHC-7}& {\bf  LHC-8}\\\hline
$pp\longrightarrow {t}{\bar t} + n-jets$ & 0.33 & 0.49 \\
$pp\longrightarrow {t}{\bar t} + b\bar{b} +  n-jets$ & 0.0014  & 0.002 \\\hline
Total   &     0.33 &      0.49\\\hline\hline
\end{tabular}
}
\caption{\sf\small Other subleading SM backgrounds (in pb) for the production process (c) followed by the decays $h\to b{\bar b}$ and $t\to b l \nu_l$ for $\sqrt{s} = 7$ and 8 TeV.}
\label{t:rbkgother}
\end{table}
%----------------------------------------------------------------------------------------%

Thus, after combining all the constraints from the Higgs observations and the top-quark decay width, one obtains the following $2\sigma$ bounds:

\begin{eqnarray}
\sqrt{\xi_{tc}^2 + \xi_{tu}^2} &\lsim & 
\begin{cases} 
0.6~(0.3)                          & {\rm for~} pp \to l^\pm l^\pm + 2b-jets + {\slashed E}_T \\
0.9~(0.9)                            & {\rm for~} pp \to l^\pm  + j + b-jet + {\slashed E}_T \\
7.5~(2.9)\times 10^{-3} & {\rm for~} pp \to l^\pm  + j + 3b-jets +  {\slashed E}_T\\
11.9~(4.8)\times 10^{-2} & {\rm for~} pp \to l^\pm  + j + b-jet + 2\gamma + {\slashed E}_T\\
4.2~(1.7)\times 10^{-3}  & {\rm for~} pp \to l^\pm  + 3j + b-jet +  {\slashed E}_T
\end{cases}
\end{eqnarray}

for the $\sqrt{s} =$ 7 TeV with 5 fb$^{-1}$ ($\sqrt{s} =$ 8 TeV with 22 fb$^{-1}$) data. 
Clearly the process $pp \to l^\pm + 3j + b-jet + {\slashed E}_T$ gives the best bound
which is about ${\cal O}(1.7 \times 10^{-3})$ using the full 8 TeV data while the 
corresponding bound for the process $pp \to l^\pm + j + b-jet + 2\gamma + {\slashed E}_T$ 
is about ${\cal O}(4.8\times 10^{-2})$ or so. A detailed list of sensitivities to all of 
these processes at individual and combined level is also presented in 
Table~\ref{t:sensitivity} for convenience. The corresponding bound on $t\to c h$ as 
reported in Ref.~\cite{Chen:2013qta} and by the ATLAS~\cite{fakejet} and CMS~\cite{cmsfla} 
experiments are about $0.1$ which is about one order of magnitude larger than the best 
bound obtained by us. This is partially due to the fact that in obtaining $t c h$ final 
state we do not restrict ourselves merely to the pair-production of $t\bar t$ unlike them. Thus in their study only diagrams of type Figs. 1 (a) and 1 (c) are relevant; in our case all three 
processes in Fig 1 occur. This translates into about $14$ times larger cross-section for us compared to $\sigma_{t\bar t} \times 
2 Br(t\to c h)$ as in their work too, the other top decays semileptonically as $t\to b l \nu_l$. Obviously, since in our case, one of the top-quarks decaying to $c h$ or $u h$, can be off-shell, it results in a further 
advantage for us in terms of reconstruction efficiencies.

\subsection{Projected sensitivities at the LHC-14}

Guided with the aforementioned bounds as obtained from various final state topologies using 
the observed 7 and 8 TeV LHC data in the previous subsection, in this subsection we provide 
estimate for the forthcoming LHC run with $\sqrt{s} = 14$ TeV. As we have noticed above, the 
best bound on the $\sqrt{\xi_{tc}^2 + \xi_{tu}^2}$ correspond to the $pp \to l^\pm + 3j + 
b-jet + {\slashed E}_T$ detection mode. We will therefore focus on this particular detection mode itself.

Using {\tt MadGraph5}, we find the bare cross-section for the production-process 
responsible for the aforementioned topology, $pp\to t c h$, at $\sqrt{s} = 14$ TeV to be 
$431.5 \xi_{tc}^2 + 441.2 \xi_{tu}^2$ pb, which is about a factor of 4 larger compared to at 8 TeV, see Table 2. This in the ideal case assuming a full 
reconstruction of the $tch$ and in the absence of any SM background yields, 
$\sqrt{\xi_{tc}^2 + \xi_{tu}^2} \lsim 0.15 \times 10^{-3}$ for an integrated luminosity 
100 fb$^{-1}$.

The cross-sections for the relevant SM processes which contribute to the background to our 
final-state are estimated to be, 6.01 (2.91) fb, 1.29 (0.59) fb, and, 1161.3 pb for $pp\to 
t{\bar b} Z (\bar t{b} Z)$, $pp\to t{\bar b} h (\bar t{b} h)$, and, $pp \to t{\bar t} + n 
j; n \leq 3$ respectively. This, using the exact same cuts as mentioned in the previous 
subsection, and, the demand that the two jets reconstruct to a Higgs-Boson, and, the only 
lepton in our final state reconstruct to a top-quark when paired with the b-jet and the 
missing transverse energy, in our final state, translates into a net SM background of 
$2.18$ fb for the final state under consideration. Thus from our complete analysis, we 
obtain $\sqrt{\xi_{tc}^2 + \xi_{tu}^2} \lsim 1.39~(1.06) \times 10^{-3}$ for $\int{\cal L} 
dt = 100~(300)$ fb$^{-1}$. It is to be noted that this is better only by a factor of about 1.6 
compared to the bound obtained using the 8 TeV data for the same signatures, while one 
naively expect it to be better by a factor of four. The reason behind this is that for 7 
and 8 TeV data, the SM background was not so significant and therefore the criterion that 
the number of observed signal events should not exceed one turn out to be stronger while 
for 14 TeV case it was actually $S/\sqrt{B}$ criterion which ruled over the former in 
estimating projected LHC sensitivities on the FCNC couplings.

%%%%%%%%%%%%%%%%%%%%%%%%%%%%%%%%%%%%%%%%%%%%%%%%%%%%%%%%%%%%%%%%%%%%%%%%%%%%%%%%%%%%%%%%%%

\section{\label{flh:concl} Conclusions}

We have studied the flavor-changing couplings of the Higgs boson with the top-quark in 
light of the Higgs discovery, using the data at the LHC, for both $\sqrt{s}= 7$ TeV and 8 
TeV, along with the Tevatron measurements of the total decay width of the top-quark.  In 
order to obtain better bounds to these processes we studied the same-sign top pair 
production, $pp \to tt ({\bar t }{\bar t})$, single top production in association with a 
light parton jet, $pp \to t {\bar j} (\bar {t} j)$, and the process where the single 
top-quark is produced in association with a Higgs and a light parton jet, $pp \to t {\bar 
j} h ( \bar {t} j h)$. In our study, we found that the process $pp \to t {\bar j} h ({\bar 
{t} j h})$ can be extremely useful in providing stronger bounds on the flavor-changing 
couplings of the order of $1.7\times 10^{-3}$, particularly in case of the latter process with the 
Higgs-boson decays into a pair of jets and the (anti)top-quark decays semileptonically 
into a b-jet, a lepton and an invisible neutrino using the full data at $\sqrt{s} = $8 
TeV. To put things in perspective, we mention in passing that these constraints are 
significantly better than the one obtained from top decays, following pair production of 
tops, and also single top production~\cite{Harnik:2012pb}. Our sensitivities on $\sqrt{\xi_{tc}^2 + \xi_{tu}^2}$ are complementary to the 
one obtained by the low energy experiments through the D0-oscillations~\cite{Harnik:2012pb} on the product  $|\xi_{tc} \xi_{tu}|$ as listed in Table 3. Note that the latter product alone is 
not sufficient particularly when one of the FCNC-couplings becomes very small or is simply zero. Our estimates for the 14 TeV 
LHC suggest that the sensitivities can be improved just a bit more to $1.1 \times 
10^{-3}$ with 300 fb$^{-1}$ data.

We also note that although the sensitivities on FCNCs are better by a factor of about 1.8 
for the process (c) in the $h \to 2~jets$ detection mode compared to $h \to b\bar{b}$, due 
to its relatively clean signature the latter can still be quite useful to constrain the 
FCNCs. This, therefore confirms that it is highly desirable to tag b's for achieving 
better sensitivities on the FCNC couplings considered in this paper. Note also that in 
principle these can be further improved by a factor of two or so provided the produced 
top-quark and the Higgs-boson could be reconstructed fully. Therefore it is worthwhile 
for the ATLAS and CMS collaborations to consider possible ways of full reconstruction to 
further constrain the crucially important flavor-changing couplings of the Higgs-boson.

%%%%%%%%%%%%%%%%%%
\begin{table}[htb]
\centering
\resizebox{16.cm}{!}
{
\begin{tabular}{| l | l | l | l |}\hline\hline
\emph{\bf S. No.}&\emph{\bf Observable}& \multicolumn{2}{|c|}{\bf $2\sigma$ sensitivity for $\sqrt{\xi_{tc}^2 + \xi_{tu}^2}$}\\\hline
(1) & $\Gamma_t$                                         & \multicolumn{2}{|c|}{$1.3$}\\\hline
(2) & ${\cal R}_{gg\gamma\gamma}$       & \multicolumn{2}{|c|}{$1.7$}\\
(3) & ${\cal R}_{gg2l2\nu}$                         & \multicolumn{2}{|c|}{$3.3$}\\
(4) & ${\cal R}_{gg4l}$                                 & \multicolumn{2}{|c|}{$1.8$}\\
(5) & $(2) + (3) + (4)$                                   & \multicolumn{2}{|c|}{$1.2$}\\\hline
\hline
\multicolumn{4}{|c|}{\bf LHC-specific}\\\hline
&           & \emph{$\sqrt{s} = 7$ TeV, $\int{\cal L} dt = 5$ fb$^{-1}$}  & \emph{$\sqrt{s} = 8$ TeV, $\int{\cal L} dt = 22$ fb$^{-1}$} \\\hline
(6)&$pp\longrightarrow t t ( {\bar t} {\bar t})$ & 0.3& 0.17 \\
(7) & $pp\longrightarrow t j ( {\bar t} j)$ & 2 & 1.3 \\
(8) & $pp\longrightarrow t j h ({\bar t} j h)$ & $1.6\times 10^{-3}$ & $0.6\times 10^{-3}$ \\\hline
(9) & $pp\longrightarrow t t ({\bar t} {\bar t}), t\to b l \nu_l$ & 0.6 & 0.3 \\
(10) & $pp\longrightarrow t j ( {\bar t} j), t\to b l \nu_l$ & 3.2 & 1.9 \\
(11) & $pp\longrightarrow t j h ( {\bar t} j h), t\to b l \nu_l, h\to b{\bar b}$ & $7.5\times 10^{-3}$ & $2.9\times 10^{-3}$ \\
(12) & $pp\longrightarrow t j h ({\bar t} j h), t\to b l \nu_l, h\to \gamma\gamma$ & $11.9\times 10^{-2}$ & $4.8\times 10^{-2}$ \\
(13) & $pp\longrightarrow t j h ({\bar t} j h), t\to b l \nu_l, h\to j j$ & $4.2\times 10^{-3}$ & $1.7\times 10^{-3}$ \\\hline
(14) & $(1) + (5) + (6)$ & 0.3 & 0.2 \\
(15) & $(1) + (5) + (7)$ & 0.9 & 0.9 \\
(16) & $(1) + (5) + (8)$ & $1.6\times 10^{-3}$ & $0.6\times 10^{-3}$ \\\hline
(17) & $(1) + (5) + (9)$ & 0.6 & 0.3 \\
(18) & $(1) + (5) + (10)$ & 0.9 & 0.9 \\
(19) & $(1) + (5) + (11)$ & $7.5\times 10^{-3}$ & $2.9\times 10^{-3}$ \\
(20) & $(1) + (5) + (12)$ & $11.9\times 10^{-2}$ & $4.8\times 10^{-2}$ \\
(21) & $(1) + (5) + (13)$ & $4.2\times 10^{-3}$ & $1.7\times 10^{-3}$ \\
\hline
\hline
\end{tabular}
}
\caption{\sf \small Upper bounds on the ${|\xi_{tc}^2 + \xi_{tu}^2|}^{1/2}$ at the $2\sigma$ level from various observations.}
\label{t:sensitivity}
\end{table}
%%%%%%%%%%%%%%%%%%

%%%%%%%%%%%%%%%%%%%%%%%%%%%%%%%%%%%%%%%%%%%%%%%%%%%%%%%%%%%%%%%%%%%%%%%%%%%%%%%%%%%%%%%%%%
\section*{\label{flh:ack} Acknowledgements}
The work of D.~A. and A.~S. are supported in part by US DOE grant Nos. DE-FG02-94ER40817 (ISU) and DE-AC02-98CH10886 (BNL). The work of S.~K.~G. was supported in part by the {\em ARC Centre of 
Excellence for Particle Physics at the Tera-scale}. The use of Monash University Sun Grid, a high-performance computing facility, is gratefully acknowledged. SKG would also like to thank the 
Regional Centre for Accelerator Based Particle Physics(RECAPP) at Harish-Chandra Research Institute, Allahabad, India for providing local hospitalities during final phase of the work.

%%%%%%%%%%%%%%%%%%%%%%%%%%%%%%%%%%%%%%%%%%%%%%%%%%%%%%%%%%%%%%%%%%%%%%%%%%%%%%%%%%%%%%%%%%

\section*{\label{flh:app} Appendix}
\appendix
\section{\label{flh:apptj} Composition of $pp\longrightarrow t {\bar q}_u ( {\bar t} q_u)$ where $q_u = \{u, c\}$}
\subsection{For ${\sqrt s} = 7$ TeV}
\begin{eqnarray}
%\resizebox{12.5cm}{!}
%\resizebox{.9\hsize}{!}{
\sigma_{pp \to t c + t {\bar c}} &=&  \sigma_{cctc} + \sigma_{c{\bar c}t{\bar c}} + \sigma_{uctc} + \sigma_{u{\bar c}t{\bar c}} + {\cal P}_{b\to c} \left(\sigma_{bctb} + \sigma_{c {\bar b}t{\bar b}} + \sigma_{butb} + \sigma_{u {\bar b}t{\bar b}}\right) +  \sigma_{b{\bar b}t{\bar c}} + \sigma_{ggt{\bar c}}\nonumber\\
%&=&  (0.1+ 0.13)\xi^2_{tc} + (0.63+ 0.63) \xi^2_{tu}+2 {\cal P}_{b\to q_u} \left(0.24 \xi^2_{tc} + 3.48 \xi^2_{tu}\right) + \left(0.26 + 20.34\right)\xi^2_{tc}  {~\rm fb} \nonumber\\
&=& (20.83 + 0.48 P_{b\to q_u}) \xi^2_{tc} + (1.26 + 6.96 P_{b\to q_u}) \xi^2_{tu}  {~\rm fb} \nonumber\\
&=& 21.03 \xi^2_{tc} +  4.18 \xi^2_{tu}   {~\rm fb} % \hspace{-2in}
\end{eqnarray}
\begin{eqnarray}
\sigma_{pp \to {\bar t} c + {\bar t} {\bar c}} &=&  \sigma_{c{\bar c}{\bar t}c} + \sigma_{{\bar c}{\bar c}{\bar t}{\bar c}} + \sigma_{{\bar u}c{\bar t}c} + \sigma_{{\bar u}{\bar c}{\bar t}{\bar c}} + {\cal P}_{b\to c} \left(\sigma_{b{\bar c}{\bar t}b} + \sigma_{{\bar c} {\bar b}{\bar t}{\bar b}} + \sigma_{b{\bar u}{\bar t}b} + \sigma_{{\bar u} {\bar b}{\bar t}{\bar b}}\right) + \sigma_{b{\bar b}{\bar t}c}  + \sigma_{gg{\bar t}c}
\nonumber\\
%&=&  (0.13+ 0.1)\xi^2_{tc} + (0.1+0.1)\xi^2_{tu} +2 {\cal P}_{b\to q_u} \left(0.24 \xi^2_{tc} + 0.51 \xi^2_{tu}\right) + \left(0.26 + 20.34\right)\xi^2_{tc}   {~\rm fb} \nonumber\\
&=& (20.83 + 0.48 P_{b\to q_u}) \xi^2_{tc} + (0.2 + 1.02 P_{b\to q_u}) \xi^2_{tu}   {~\rm fb} \nonumber\\
&=& 21.03 \xi^2_{tc} +  0.63 \xi^2_{tu}   {~\rm fb}  % \hspace{2in}
\end{eqnarray}
\begin{eqnarray}
\sigma_{pp \to t u + t {\bar u}} &=&  \sigma_{cutu} + \sigma_{c{\bar u}t{\bar u}} + \sigma_{uutu} + \sigma_{u{\bar u}t{\bar u}} + {\cal P}_{b\to u}  \left(\sigma_{bctb} + \sigma_{c {\bar b}t{\bar b}} + \sigma_{butb} + \sigma_{u {\bar b}t{\bar b}}\right) + \sigma_{b{\bar b}t{\bar u}} +\sigma_{ggt{\bar u}}\nonumber\\
%&=&  \{(3.42+ 0.49)\xi^2_{tc} + (34.4+13.2)\xi^2_{tu}\}\times 10^{-6} +2 {\cal P}_{b\to q_u} \left(0.24 \xi^2_{tc} + 3.48 \xi^2_{tu}\right) + \left(0.26 + 20.34\right)\xi^2_{tu}   {~\rm fb} \nonumber\\
&=& (3.91\times 10^{-6} + 0.48 P_{b\to q_u}) \xi^2_{tc} + (20.6 + 6.96 P_{b\to q_u}) \xi^2_{tu}  {~\rm fb}  \nonumber\\
&=& 0.2\xi^2_{tc} + 23.52 \xi^2_{tu}   {~\rm fb} %\hspace{-2in}
\end{eqnarray}
\begin{eqnarray}
\sigma_{pp \to {\bar t} u + t {\bar u}} &=&  \sigma_{u{\bar c}{\bar t}u} + \sigma_{{\bar u}{\bar c}{\bar t}{\bar u}} + \sigma_{{\bar u}u{\bar t}u} + \sigma_{{\bar u}{\bar u}{\bar t}{\bar u}} + {\cal P}_{b\to u} \left(\sigma_{b{\bar c}{\bar t}b} + \sigma_{{\bar c} {\bar b}{\bar t}{\bar b}} + \sigma_{b{\bar u}{\bar t}b} + \sigma_{{\bar u} {\bar b}{\bar t}{\bar b}}\right) + \sigma_{b{\bar b}{\bar t}u} + \sigma_{gg{\bar t}u}\nonumber\\
%&=&  \{(3.42+ 0.49)\xi^2_{tc} + (14.6+2.04)\xi^2_{tu}\}\times 10^{-6} +2 {\cal P}_{b\to q_u} \left(0.24 \xi^2_{tc} + 0.51 \xi^2_{tu}\right) + \left(0.26 + 20.34\right)\xi^2_{tu}   {~\rm fb} \nonumber\\
&=& (3.91\times 10^{-6} + 0.48 P_{b\to q_u}) \xi^2_{tc} + (20.6 + 1.02 P_{b\to q_u}) \xi^2_{tu}  {~\rm fb}  \nonumber\\
&=& 0.2 \xi^2_{tc} +21.03 \xi^2_{tu}   {~\rm fb}  %\hspace{-2in}
\end{eqnarray}

Here we have assumed $P_{b\to q_u} = 0.42$.

\subsection{For ${\sqrt s} = 8$ TeV}
\begin{eqnarray}
\sigma_{pp \to t c + t {\bar c}} &=&  \sigma_{cctc} + \sigma_{c{\bar c}t{\bar c}} + \sigma_{uctc} + \sigma_{u{\bar c}t{\bar c}} + {\cal P}_{b\to c} \left(\sigma_{bctb} + \sigma_{c {\bar b}t{\bar b}} + \sigma_{butb} + \sigma_{u {\bar b}t{\bar b}}\right) +  \sigma_{b{\bar b}t{\bar c}} + \sigma_{ggt{\bar c}}\nonumber\\
%&=&  (0.14+ 0.16)\xi^2_{tc} + (0.77+ 0.77) \xi^2_{tu}+2 {\cal P}_{b\to q_u} \left(0.34 \xi^2_{tc} + 4.24 \xi^2_{tu}\right) + \left(0.37 + 28.89\right)\xi^2_{tc}   {~\rm fb} \nonumber\\
&=& (29.56 + 0.68 P_{b\to q_u}) \xi^2_{tc} + (1.54+ 8.48 P_{b\to q_u}) \xi^2_{tu}   {~\rm fb} \nonumber\\
&=& 29.85 \xi^2_{tc} + 5.1 \xi^2_{tu}   {~\rm fb} 
\end{eqnarray}
\begin{eqnarray}
\sigma_{pp \to {\bar t} c + {\bar t} {\bar c}} &=&  \sigma_{c{\bar c}{\bar t}c} + \sigma_{{\bar c}{\bar c}{\bar t}{\bar c}} + \sigma_{{\bar u}c{\bar t}c} + \sigma_{{\bar u}{\bar c}{\bar t}{\bar c}} + {\cal P}_{b\to c} \left(\sigma_{b{\bar c}{\bar t}b} + \sigma_{{\bar c} {\bar b}{\bar t}{\bar b}} + \sigma_{b{\bar u}{\bar t}b} + \sigma_{{\bar u} {\bar b}{\bar t}{\bar b}}\right) + \sigma_{b{\bar b}{\bar t}c}  + \sigma_{gg{\bar t}c}
\nonumber\\
%&=&  (0.16+ 0.14)\xi^2_{tc} + (0.14+0.14)\xi^2_{tu} +2 {\cal P}_{b\to q_u} \left(0.34 \xi^2_{tc} + 0.7 \xi^2_{tu}\right) + \left(0.37 + 28.89\right)\xi^2_{tc}   {~\rm fb} \nonumber\\
&=& (29.56 + 0.68 P_{b\to q_u}) \xi^2_{tc} + (0.28 + 1.4 P_{b\to q_u}) \xi^2_{tu}   {~\rm fb} \nonumber\\
&=& 29.85 \xi^2_{tc} +  0.34 \xi^2_{tu}   {~\rm fb} 
\end{eqnarray}
\begin{eqnarray}
\sigma_{pp \to t u + t {\bar u}} &=&  \sigma_{cutu} + \sigma_{c{\bar u}t{\bar u}} + \sigma_{uutu} + \sigma_{u{\bar u}t{\bar u}} + {\cal P}_{b\to u}  \left(\sigma_{bctb} + \sigma_{c {\bar b}t{\bar b}} + \sigma_{butb} + \sigma_{u {\bar b}t{\bar b}}\right) + \sigma_{b{\bar b}t{\bar u}} +\sigma_{ggt{\bar u}}\nonumber\\
%&=&  \{(4.23+ 0.66)\xi^2_{tc} + (37.2+15.7)\xi^2_{tu}\}\times 10^{-6} +2 {\cal P}_{b\to q_u} \left(0.33 \xi^2_{tc} + 4.24 \xi^2_{tu}\right) + \left(0.37 + 28.89\right)\xi^2_{tu}   {~\rm fb} \nonumber\\
&=& (4.89\times 10^{-6} + 0.66 P_{b\to q_u}) \xi^2_{tc} + (29.26 + 0.66 P_{b\to q_u}) \xi^2_{tu}   {~\rm fb}  \nonumber\\
&=& 0.32 \xi^2_{tc} + 29.54 \xi^2_{tu}   {~\rm fb} 
\end{eqnarray}
\begin{eqnarray}
\sigma_{pp \to {\bar t} u + t {\bar u}} &=&  \sigma_{u{\bar c}{\bar t}u} + \sigma_{{\bar u}{\bar c}{\bar t}{\bar u}} + \sigma_{{\bar u}u{\bar t}u} + \sigma_{{\bar u}{\bar u}{\bar t}{\bar u}} + {\cal P}_{b\to u} \left(\sigma_{b{\bar c}{\bar t}b} + \sigma_{{\bar c} {\bar b}{\bar t}{\bar b}} + \sigma_{b{\bar u}{\bar t}b} + \sigma_{{\bar u} {\bar b}{\bar t}{\bar b}}\right) + \sigma_{b{\bar b}{\bar t}u} + \sigma_{gg{\bar t}u}\nonumber\\
%&=&  \{(4.23+ 0.64)\xi^2_{tc} + (17.1+2.63)\xi^2_{tu}\}\times 10^{-6} +2 {\cal P}_{b\to q_u} \left(0.34 \xi^2_{tc} + 0.7 \xi^2_{tu}\right) + \left(0.37 + 28.89\right)\xi^2_{tu}   {~\rm fb} \nonumber\\
&=& (4.87\times 10^{-6} + 0.68 P_{b\to q_u}) \xi^2_{tc} + (29.26 + 1.4 P_{b\to q_u}) \xi^2_{tu}  {~\rm fb}  \nonumber\\
&=& 0.29 \xi^2_{tc} + 30.13 \xi^2_{tu}   {~\rm fb} 
\end{eqnarray}

%%%%%%%%%%%%%%%%%%%%%%%%%%%%%%%%%%%%%%%%%%%%%%%%%%%%%%%%%%%%%%%%%%%%%%%%%%%%%%%%%%%%%%%%%%
\bibliographystyle{jhep} % Referencing style
\bibliography{flh} % The name of the .bib file
%%%%%%%%%%%%%%%%%%%%%%%%%%%%%%%%%%%%%%%%%%%%%%%%%%%%%%%%%%%%%%%%%%%%%%%%%%%%%%%%%%%%%%%%%%
\end{document}